\documentclass[aps,prd,10pt,groupedaddress,twocolumn,showpacs,floatfix]{revtex4-1}
\usepackage{textcomp,gensymb}
\usepackage{hyperref}
\usepackage{graphicx}
\usepackage{amsfonts,amsmath,amssymb,bm,bbm}
\usepackage{color,soul}
\usepackage{color}
\usepackage{cancel}
\usepackage{slashed}
\usepackage[toc,page]{appendix}
\usepackage{flushend}
\usepackage{physics}
\usepackage{graphicx}
\usepackage{dcolumn}
\usepackage{float}

\usepackage{fontawesome} 

\graphicspath{{./figures/}}
\definecolor{lightgreen}{rgb}{.14,.44,.14}

\hypersetup{
    pdfnewwindow=true,      
    colorlinks=true,       
    linkcolor=blue,          
    citecolor=blue,        
    filecolor=blue,      
    urlcolor=blue        
}

\begin{document}

\title{Limiting Light Dark Matter with Luminous Hadronic Loops}

\author{Melissa Diamond}
\email{m.diamond@queensu.ca}

\author{Christopher V. Cappiello}
\email{cvc1@queensu.ca}

\author{Aaron C. Vincent}
\email{aaron.vincent@queensu.ca}

\author{Joseph Bramante}
\email{joseph.bramante@queensu.ca}
\affiliation{Department of Physics, Engineering Physics, and Astronomy, Queen's University, Kingston, Ontario, K7N 3N6, Canada,\\
The Arthur B. McDonald Canadian Astroparticle Physics Research Institute, Kingston, Ontario, K7L 3N6, Canada,\\
\textit{and} Perimeter Institute for Theoretical Physics, Waterloo, Ontario, N2L 2Y5, Canada}

\begin{abstract}
Dark matter is typically assumed not to couple to the photon at tree level. While annihilation to photons through quark loops is often considered in indirect detection searches, such loop-level effects are usually neglected in direct detection, as they are typically subdominant to tree-level dark matter-nucleus scattering. However, when dark matter is lighter than around 100 MeV, it carries so little momentum that it is difficult to detect with nuclear recoils at all.  We show that loops of low-energy hadronic states can generate an effective dark matter-photon coupling, and thus lead to scattering with electrons even in the absence of tree-level dark matter-electron scattering. For light mediators, this leads to an effective fractional electric charge which may be very strongly constrained by astrophysical observations. Current and upcoming searches for dark matter-electron scattering can thus set limits on dark matter-proton interactions down to 1 MeV and below.
\end{abstract}

\maketitle


\section{Introduction}

Although dark matter (DM) makes up most of the mass in the Universe, how (or even whether) it interacts non-gravitationally is completely unknown~\cite{Bertone:2004pz,Planck:2018vyg,Cooley:2022ufh}. Direct detection bounds on DM, as well as astrophysical and cosmological constraints, are often presented in a ``model-independent'' way, e.g. as limits on dark matter's nonrelativistic scattering cross section with protons or electrons, rather than limits on a larger set of model parameters. In such a framework, limits on dark matter's interactions with different Standard Model particles are often treated completely independently. There are of course exceptions---DM-proton and DM-neutron cross sections are often assumed to be identical to avoid isospin violation---but for example, limits on DM-electron scattering are often set assuming that DM-nucleon scattering is negligible, and vice-versa.

However, even if DM only interacts with one Standard Model particle at tree level, interactions with other particles can be generated at loop level. A classic example of such a loop calculation is Ref.~\cite{Holdom:1985ag}, which showed how fermionic loops could generate an effective electromagnetic charge. This example is commonly cited in literature on dark photons and millicharged DM~\cite{Prinz:1998ua,Davidson:2000hf,Pospelov:2008zw,McDermott:2010pa,BaBar:2014zli,Izaguirre:2015eya,Mahdawi:2018euy,Munoz:2018pzp,Emken:2019tni,Caputo:2021eaa,Chiles:2021gxk,Kahn:2021ttr,Gorghetto:2022sue,Romanenko:2023ytm}, and explicit calculations of quark-, lepton-, or W-loop-induced processes may be found in many references on indirect detection, collider searches, and, to a lesser extent, direct detection~(\cite{Bern:1997ng,Bergstrom:1997fh,Ullio:1997ke,Boudjema:2005hb,Fox:2008kb,Kopp:2009et,Bringmann:2012ez,Bell:2014tta,Bramante:2015una,Fermi-LAT:2016afa,Kumar:2016gxq,Braaten:2017dwq,Bringmann:2018lay,HESS:2018cbt,FileviezPerez:2019rcj,HAWC:2019jvm,Arina:2020udz,Rinchiuso:2020skh,Arina:2021gfn,Foster:2022nva}, see also~\cite{Kopp:2014tsa,Mohan:2019zrk}). In works focused on indirect detection or collider searches, the energies are typically high enough that QCD is perturbative, and loops of quarks can be computed directly.

In this work, we explicitly compute the DM-photon interaction induced by DM couplings to hadronic states at low energy, where the relevant degrees of freedom are not quarks, but mesons and baryons. This is in contrast with much of the literature on kinetically mixed dark photons, which treats the mixing between the photon and dark photon as a phenomenological parameter that is generated at much higher energy scales. We show that such loop-level couplings can produce detectable event rates in direct detection experiments, and/or induce a non-negligible effective charge for the DM, in a wide range of sub-GeV parameter space. We thus set new limits on sub-GeV dark matter's coupling to protons.

This paper is organized as follows. In Section~\ref{sec:loops}, we review the ideas of Ref.~\cite{Holdom:1985ag}, and discuss the types of interaction that yield nonzero results. We then introduce both our model and the effective Lagrangians used to describe dark matter's interactions with hadronic states, and compute the induced interactions with photons and electrons. In Section~\ref{sec:results}, we compute new constraints on sub-GeV DM. In Section~\ref{sec:conclusions}, we discuss the implications of our results. Detailed descriptions of the calculations performed in this work can be found in the Appendix~\ref{sec:appendixA}.

\section{Effective Dark Matter-electron Interactions}\label{sec:loops}

\begin{figure}[h]
\centering
        \includegraphics[width=\columnwidth]{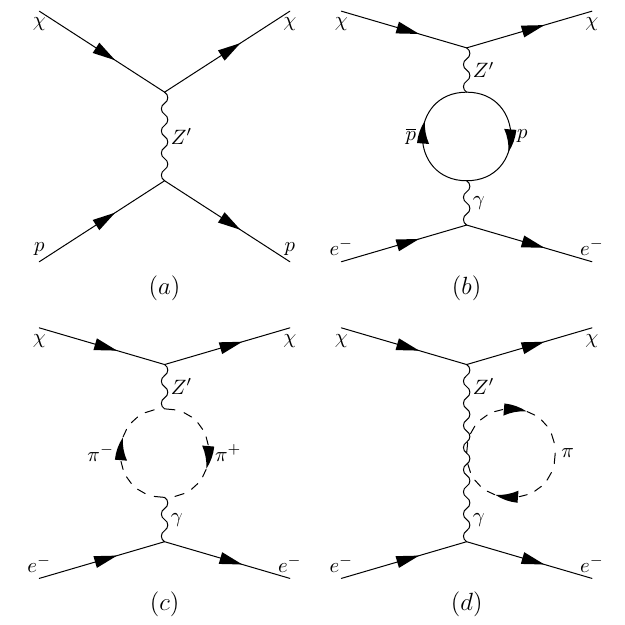}
        
    \caption{\textbf{\textit{Feynman diagrams utilized in this work.}}  The tree level diagram (a) captures DM-proton scattering.  The loop level diagrams (b)-(d) show DM-electron scattering that results from the DM-proton interaction.}
    \label{fig:loops}
\end{figure}

Suppose that DM is a fermion, and interacts with a particular charged Standard Model fermion---in this case, a proton---via a new vector mediator that we will refer to as $Z'$ (the tree level diagram is shown in Fig.~\ref{fig:loops} (a)). A proton loop then induces a mixing between the $Z'$ and the photon, as well as a DM-electron interaction, as shown in Fig.~\ref{fig:loops} (b). If $Z'$ is massless, the result is an effective charge for the DM, as derived by Ref.~\cite{Holdom:1985ag}. Even if the mediator is massive, a mixing with the photon is still generated, but the different momentum dependence means that the DM no longer behaves as a truly millicharged particle.

The proton is, of course, not the only hadronic state that can be included in loops like this. In fact, the more typical approach to such hadronic corrections would be to use loops of pions (Fig.~\ref{fig:loops} (c) and (d)), whose masses are below the QCD scale and for which the framework of chiral effective field theory (ChEFT) can readily be applied. The use of pion loops in this context is strikingly similar to their role in hadronic vacuum polarization, notably in the context of muon $g-2$ (see Ref.~\cite{Jegerlehner:2017gek} for a detailed review). 

In this work, we include both proton and meson (specifically pion and kaon) loops in order to compute the induced mixing between the $Z'$ and photon. We take inspiration from Ref.~\cite{Becher:1999he}, which showed that nucleons could be included in the framework of chiral perturbation theory while still preserving chiral power counting.

Throughout this work, we will consider only the case of a vector mediator, because for a scalar or axial vector, the diagrams shown in Fig.~\ref{fig:loops} (b)-(d) vanish (see Ref.~\cite{Kopp:2009et}). An analogous interaction between DM and electrons in the scalar case can be induced at the 2-loop level~\cite{Kopp:2009et}, or by instead mixing with the Higgs at one loop.  However, these options are suppressed compared to the one-loop mixing with the photon, so we do not consider them.

Next, we will estimate the effective DM-electron scattering cross section that results from couplings through hadronic loops.   We focus on DM masses in the range 1 MeV $\lesssim m_{\chi} \lesssim 100 $ MeV, which is difficult to probe using nuclear recoil searches, but where electron recoil searches have set some bounds~\cite{Essig:2017kqs,DAMIC:2019dcn,XENON:2019gfn,SENSEI:2020dpa,PandaX-II:2021nsg,XENON:2021qze,DarkSide:2022knj,DAMIC-M:2023gxo}. At the corresponding energies ($T_{\chi} \ll 1$ keV), quarks are confined into baryons and light mesons, and their behavior is best described using ChEFT.  Hadronic loops are dominated by pions and kaons, the light pseudoscalar mesons.  Proton loops will contribute as well, and may dominate depending on the underlying theory.  We start with description of the DM quark interaction and use this to build a consistent description of the DM-proton and DM-meson coupling.  These will then be used to estimate and compare the tree level DM-proton cross section with the 1-loop DM-electron cross section.  

Interactions between a dark fermion and quarks through a vector mediator  can be described by
\begin{equation}
    \label{eq:quarkcouple}\mathcal{L}\supset\sum_q\alpha_q Z'_{\mu}q\gamma^{\mu}\bar{q}+g_{\chi}Z'_{\mu}\chi\gamma^{\mu}\bar{\chi}\,,
\end{equation}
where $\chi$ is the massive DM particle, $q$ are the quarks, $\alpha_q$ is the coupling of each quark specie to $Z'_{\mu}$, $g_{\chi}$ is the coupling between the $Z'$ and the DM. 

We may define a resulting effective proton interaction~\cite{agrawal2010classification}: 
\begin{equation}
    \mathcal{L}\supset (2\alpha_u+\alpha_d) Z'_{\mu}p\gamma^{\mu}\bar{p}\,.
\end{equation}
The meson interaction terms can be derived from the ChEFT Lagrangian.  The relevant lowest order interaction term in the ChEFT Lagrangian is \cite{Ecker:1988te}
\begin{equation}
\label{eq:chiralL}
    L\supset \frac{F^2}{4}Tr\left(D_{\mu}U D^{\mu}U^{\dag}\right)\,,
\end{equation}
where $U= e^{\frac{i}{F}\pi}$ contains the light meson octet 
\begin{equation}
    \pi= 
    \begin{pmatrix}
    \pi^0+\frac{\eta_8}{\sqrt{3}}&\sqrt{2}\pi^+ &\sqrt{2}K^+ \\ \sqrt{2}\pi^-&-\pi^0+\frac{\eta_8}{\sqrt{3}} &\sqrt{2}K^0\\
    \sqrt{2}K^-&\sqrt{2}\bar{K}^0 & -\frac{2}{\sqrt{3}}\eta_8
    \end{pmatrix}\,.
    \end{equation}
$F$ is the pion decay constant.  Interactions with external vector fields (namely the photon or $Z'$) are captured in the derivative terms
\begin{equation}
\begin{split}
    &D_{\mu}U =\partial_{\mu}U-iv_{\mu} U+iUv_{\mu}\,,\\
    &D_{\mu}U^{\dag}=\partial_{\mu}U^{\dag}+iU^{\dag}v_{\mu}-iv_{\mu} U^{\dag}\,,\\
\end{split}
\end{equation}
where
$v_{\mu}=Z'_{\mu}\text{diag}(\alpha_u,\alpha_d,\alpha_s)$ is a matrix which represent an external vector interaction emerging from interactions of the form depicted in \eqref{eq:quarkcouple}.  We can include electromagnetic interactions here as well by including a term of the form $v_{\mu} = e A_{\mu}\text{diag}(\frac{2}{3},-\frac{1}{3},-\frac{1}{3})$.  

Expanding out the chiral Lagrangian \eqref{eq:chiralL} gives the following interaction terms between light mesons, photons, and $Z'$.
\begin{equation}
\begin{split}
    \mathcal{L}\supset &i(\alpha_u-\alpha_d)Z'_{\mu}(\pi^-\partial^{\mu}\pi^+ - \pi^+\partial_\mu\pi^-)\\+&i(\alpha_u-\alpha_s)Z'_{\mu}(K^-\partial^{\mu}K^+ - K^+\partial^{\mu} K^-)\\
    +&ieA_{\mu}(\pi^-\partial^{\mu}\pi^+ - \pi^+\partial_\mu\pi^-)\\+&ieA_{\mu}(K^-\partial^{\mu}K^+ - K^+\partial{\mu} K^-)\\
    +&2 e(\alpha_u-\alpha_d)Z'_{\mu}A^{\mu}\pi^+\pi^-\\
    +&2 e(\alpha_u-\alpha_s)Z'_{\mu}A^{\mu}K^+K^-\,.\\
\end{split}
\end{equation}

Using these interaction terms we calculate the $\chi p$ scattering cross section at tree level and the $\chi e$ scattering cross section at loop level based on the diagrams shown in Fig.~\ref{fig:loops}.  These calculations are shown in more detail in the Appendix~\ref{sec:appendixA}.

At tree level and low momentum exchange 
\begin{equation}
\label{protontree}
    \frac{d\sigma_{\chi p}}{d\Omega} = \frac{g_{\chi}^2 (2\alpha_u+\alpha_d)^2 \mu_{\chi p}^2}{4\pi^2 (m_Z^2-t)^2 }\,,
\end{equation}
where $m_{Z'}$is the $Z'$ mass. $\mu_{a b} = {m_a m_b}/({m_a+m_b})$ represents the reduced mass of two particles with masses $m_a$ and $m_b$. 

The loop terms together give a cross section of
\begin{equation}
    \label{eq:heavyconv}
    \frac{d\sigma_{\chi e}}{d\Omega} =
   \frac{d\sigma_{\chi p}}{d\Omega} \frac{e^2}{2304 \pi^4 (2 \alpha_u+\alpha_d)^2}  \left(\frac{\mu_{\chi e}}{\mu_{\chi p}}\right)^2c_{loop}^2\,,
\end{equation}
where 

\begin{equation}
\begin{split}
    c_{loop} = \,\,& 4(2\alpha_u+\alpha_d)\ln\left(\frac{4 \pi e^{-\gamma_E}\mu^2}{m_{p}^2}\right)\\
    &+(\alpha_u-\alpha_d)\ln\left(\frac{4 \pi e^{-\gamma_E}\mu^2}{m_{\pi}^2}\right)\\
    &+(\alpha_u-\alpha_s)\ln\left(\frac{4 \pi e^{-\gamma_E}\mu^2}{m_{K}^2}\right)\,.
\end{split}    
\end{equation}
The first term in $c_{loop}$ comes from the proton loop as shown in Fig.~\ref{fig:loops} (b), the second term comes from a pion loop and the third from the kaon loop. The meson loop terms get contributions from diagrams of the form depicted in Fig.~\ref{fig:loops} (c) and (d).  Loop divergences are contained in the log terms, which depend on the mass of the particles traveling through the loops and a cutoff term that is logarithmic in $\mu$.  We follow Ref.~\cite{Becher:1999he} and set $\mu=m_p$.   We use  $m_{\pi}=140$ MeV, $m_K = 494$ MeV \cite{Workman:2022ynf}, and $m_p = 938$ MeV. $\gamma_E = 0.577$ is the Euler Mascheroni constant. 

In the case of a heavy mediator, one can integrate over scattering angles and give a relative total cross section of
\begin{equation}
    \sigma_{\chi e} = \sigma_{\chi p} \frac{e^2}{2304 \pi^4 (2 \alpha_u+\alpha_d)^2}  \left(\frac{\mu_{\chi e}}{\mu_{\chi p}}\right)^2c_{loop}^2\,.
    \label{eq:approxheavyconv}
   \end{equation}
The factor $\frac{e^2}{2304 \pi^4 }  \approx4\times 10^{-7}$.  While this suppression seems substantial, the effective cross sections are large enough that planned and currently running electron recoil detectors should be able to observe or rule out DM that is difficult to observe using traditional nuclear recoil detectors. 

For a light mediator, the integral of ${d\sigma_{\chi e}}/{d\Omega}$ and $
   {d\sigma_{\chi p}}/{d\Omega}$ over the scattering angle diverges.  So we instead report the ratio between $\bar{\sigma}_{\chi p}$ and $\bar{\sigma}_{\chi e}$, where
   \begin{equation}
       \bar{\sigma} \equiv 2\frac{d\sigma}{d\cos{\theta}}\left(\frac{q}{q_{ref}}\right)^4. 
   \end{equation}
Here $q$ represents the momentum exchanged between scattered particles, and $q_{ref}$ is a reference momentum, usually taken to be $\sim\alpha m_e$~\cite{Emken:2019tni}.  $\bar{\sigma}$ is Lorentz-invariant and typically used when discussing constraints on light mediator scattering.  The resulting relation between light mediator cross sections is
\begin{equation}
    \bar{\sigma}_{\chi e} =
   \bar{\sigma}_{\chi p} \frac{e^2}{2304 \pi^4 (2 \alpha_u+\alpha_d)^2}  \left(\frac{\mu_{\chi e}}{\mu_{\chi p}}\right)^2c_{loop}^2\,.
\end{equation}

 Hence, the ratio between the proton and electron cross sections does not depend on mass of the vector mediator.  As the terms shared between proton and electron scattering diagrams divide out, the ratio between the proton and electron cross section also does not depend on the spin of the DM, or the Lorentz structure of its interaction with the vector mediator.

\section{Results}\label{sec:results}

In direct detection literature, the interactions between DM and individual quarks that generate the DM-nucleon cross section are typically left unspecified. Because we also include meson loops, recasting these limits requires a concrete choice for the couplings to individual quarks. In this Section, we report our results for the case $\alpha_u=-\alpha_d$ and $\alpha_s=0$, i.e. the case where the $Z'$ couples to isospin. Results for an alternative case, $\alpha_u=\alpha_d=\alpha_s$, are shown in the Appendix~\ref{sec:appendixB}.  The limits between these two cases differ by a factor of $\sim 8$.  We do not expect different choices of couplings to weaken the limits much beyond this range without fine tuning.

For our constraints on the DM-proton cross section using electron-recoil searches, we focus on SENSEI~\cite{SENSEI:2020dpa}, which has reported some of the strongest limits on DM-electron scattering while being only $\sim$100 meters underground (just under 300 meters water equivalent, or m.w.e.), presenting a lower overburden than most direct detection experiments. Our limits result from rescaling the reported limits of Ref.~\cite{SENSEI:2020dpa} by the ratio of the DM-proton and loop-induced DM-electron cross sections. In the same way, we also recast projections for the DAMIC-M experiment~\cite{Battaglieri:2017aum,Settimo:2018qcm,Settimo:2020cbq}, which has recently released its first results~\cite{DAMIC-M:2023gxo}, as well as the upcoming Oscura experiment~\cite{2022arXiv220210518A}.  

\begin{figure}
    \centering
    \includegraphics[width=\columnwidth]{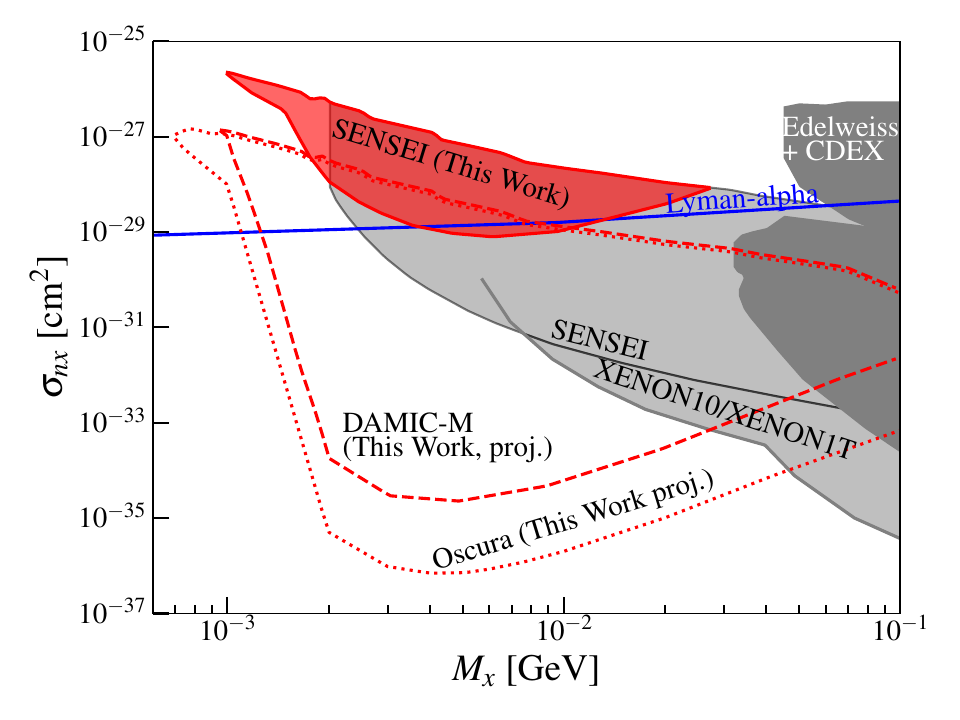}
    \caption {\textit{\textbf{Limits on the dark matter-nucleon cross section due to their loop-induced coupling to electrons. Interactions via a heavy vector mediator}} with $\alpha_u=-\alpha_d$ and $\alpha_s=0$.   Our recasting of constraints from SENSEI ~\cite{SENSEI:2020dpa} is shown in red, while the regions outlined in red dashed and dotted lines will be accessible to DAMIC-M ~\cite{Battaglieri:2017aum,Settimo:2018qcm,Settimo:2020cbq} and Oscura ~\cite{2022arXiv220210518A} respectively.  Existing detector constrains From Migdal effect searches at SENSEI \cite{SENSEI:2020dpa,Berghaus:2022pbu}, XENON10/1T  (\cite{Essig:2019xkx}, as shown as in Ref.~\cite{Berghaus:2022pbu}), CDEX \cite{CDEX:2019hzn,CDEX:2021cll} and EDELWEISS \cite{EDELWEISS:2019vjv,2022PhRvD.106f2004A} are shown in gray, while Lyman alpha constraints \cite{Rogers:2021byl} are shown in blue.}
    \label{fig:heavy}
\end{figure}

Figure~\ref{fig:heavy} shows limits and projected sensitivities based on effective loop interactions,  for a heavy $Z'$, compared to existing limits from direct detection and cosmology. Our limit constrains DM masses from about 1 MeV to 30 MeV, and at the lowest masses is comparable to the strongest existing Migdal effect limit, from SENSEI~\cite{SENSEI:2020dpa,Berghaus:2022pbu}. It is also competitive with the strongest cosmological bounds, which come from Lyman-alpha observations~\cite{Rogers:2021byl}. The projected sensitivities reach cross sections of 10$^{-36}$ cm$^2$ for masses of a few MeV, orders of magnitude better than existing Migdal effect searches and Lyman-alpha bounds. For comparison, Ref.~\cite{2022arXiv220210518A} shows projections for future Migdal effect searches from the same detectors, but these projections only extend down to 20 MeV. We also show Migdal effect limits from XENON10 and XENON1T~\cite{Essig:2019xkx}, CDEX~\cite{CDEX:2019hzn,CDEX:2021cll}, and EDELWEISS~\cite{EDELWEISS:2019vjv,2022PhRvD.106f2004A}, which provide leading constraints at larger masses.

At the large cross sections we consider, DM may be stopped in the Earth before reaching the detector. Even though the signal we use to set limits is scattering with electrons, attenuation will be dominated by scattering with nuclei due to the much larger cross section. To account for attenuation in the Earth, we use the ceilings computed for SENSEI in Ref.~\cite{Emken:2019tni}. These ceiling calculations are also dominated by nuclear scattering, but were computed in a dark photon model, so we rescale them by a factor of 4 to account for the scattering with neutrons (assuming typical spin-independent scattering). For DAMIC-M and Oscura, we use the same ceiling, lowered by factors of 17 and 20, respectively, to simulate the overburdens at Laboratoire Souterrain de Modane (4800 m.w.e.)~\cite{Piquemal:2012fs} and SNOLAB (6000 m.w.e.)~\cite{Duncan:2010zz}. 

\begin{figure}
    \centering
    \includegraphics[width=\columnwidth]{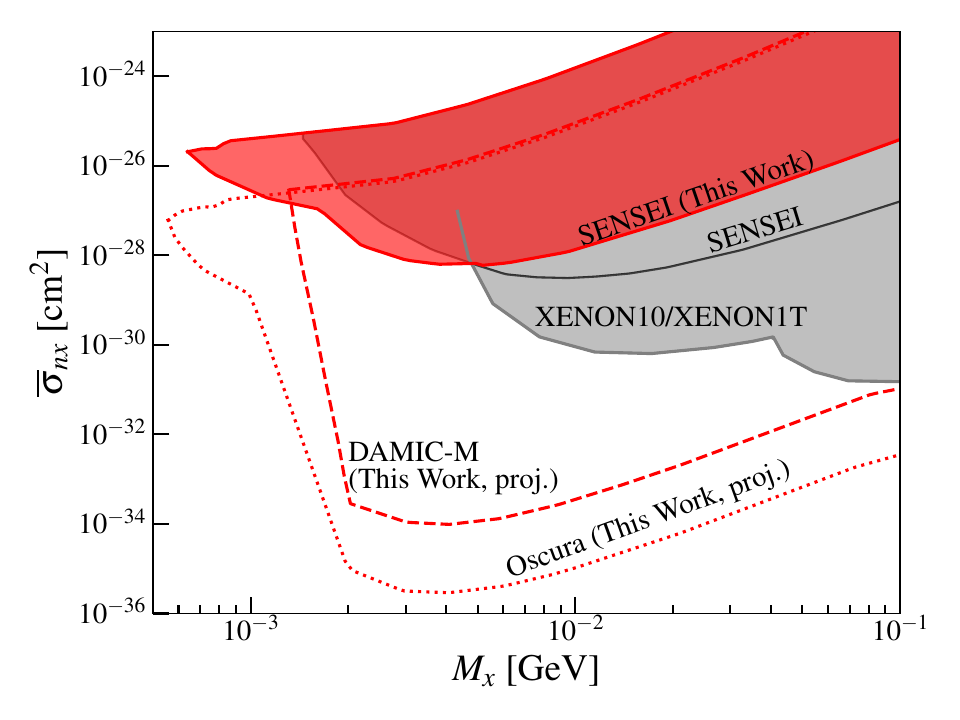}
    \caption {\textbf{\textit{Limits on the dark matter-nucleon cross section due to their loop-induced coupling to electrons. Interactions via a light vector mediator}} with $\alpha_u=-\alpha_d$ and $\alpha_s=0$. Our results for SENSEI ~\cite{SENSEI:2020dpa}  are highlighted in red, and the expected reach of DAMIC-M ~\cite{Battaglieri:2017aum,Settimo:2018qcm,Settimo:2020cbq} and Oscura ~\cite{2022arXiv220210518A} are outlined by the red dashed and dotted lines respectively. Constraints from Migdal effect searches at SENSEI \cite{SENSEI:2020dpa,Berghaus:2022pbu}, and XENON10/1T (\cite{Essig:2019xkx}, as shown in Ref.~\cite{Berghaus:2022pbu}) are shown in gray.}
    \label{fig:light}
\end{figure}

Figure~\ref{fig:light} shows our limits and projected sensitivities for a light $Z'$, compared to existing limits from direct detection (cosmological bounds exist on scattering via light mediators, e.g.~\cite{Kovetz:2018zan,Slatyer:2018aqg} but only at higher cross section). As mentioned above, in the case of a massless mediator, the total cross section diverges, and limits are typically reported in terms of a reference cross section $\overline{\sigma}$. We follow the parametrization of Ref.~\cite{Emken:2019tni} (see also the Appendix ~\ref{sec:appendixA} for additional discussion).  Also in the light $Z'$ case, the scattering is typically softer, making attenuation less of an issue, so we can constrain a much wider range of parameter space. At large DM masses, Migdal effect bounds from SENSEI and XENON10/XENON1T are stronger than our bounds. However, our limits are stronger than existing Migdal effect limits for masses up to 5 MeV, and extend down well below 1 MeV. In our projections, we show that DAMIC-M and Oscura can again probe cross sections in the range 10$^{-34}$--10$^{-36}$ cm$^2$, competitive with the Migdal effect projections from Ref.~\cite{2022arXiv220210518A} and surpassing them for masses below $\sim$10 MeV if one were to extrapolate those projections to lower mass. We again show direct detection constraints from SENSEI, XENON10, and XENON1T for comparison. It deserves mention that models of new light $Z'$ mediators coupled to baryon currents will also induce SM anomalies that can be constrained through their contribution to rare meson decays \cite{Dror:2017nsg,Dror:2018wfl}.

Finally, we note that an effectively massless $Z'$ produces an effective fractional electric charge (or ``millicharge'') for the DM. We can compute the effective charge induced by loops of hadronic states, and recast limits on millicharged DM as limits on DM-proton interactions via an effectively massless vector. We report our results in terms of $\epsilon$, the DM charge in units of the electron charge, i.e. $q_{DM} = \epsilon e$. 

\begin{figure}[!ht]
    \centering
    \hspace{-8mm}\includegraphics[width=1.1\columnwidth]{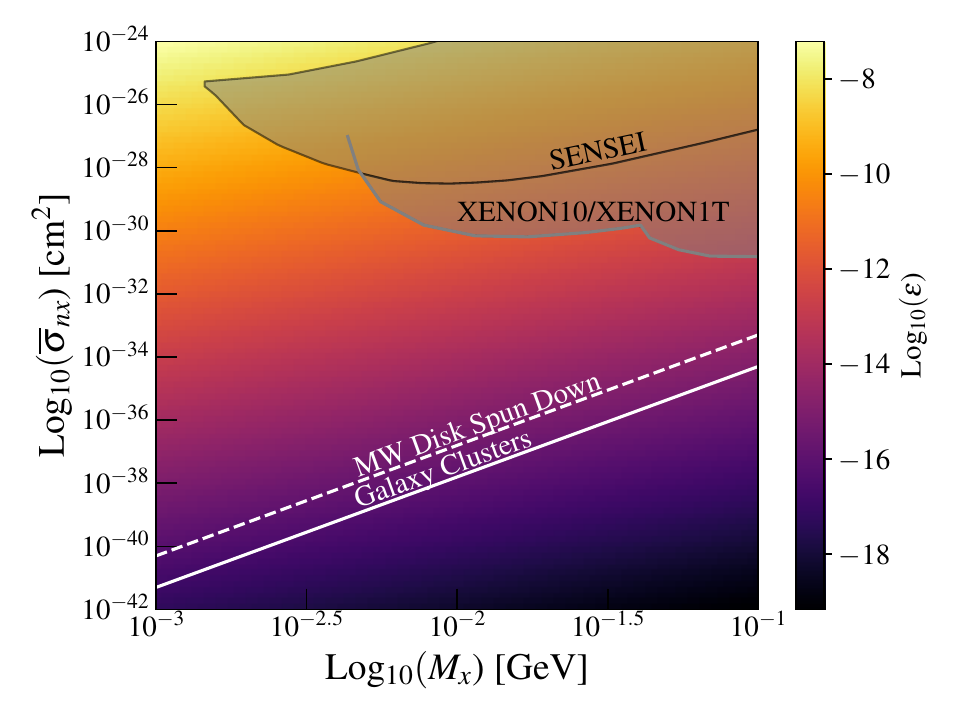}
    \caption{\textbf{\textit{Effective electric charge resulting from hadronic loop interactions}} assuming $\alpha_u=-\alpha_d$ and $\alpha_s=0$. The white lines are astrophysical constaints on millicharged DM: above the solid line, cluster magnetic fields would noticeably alter the density profile of galaxy clusters~\cite{Kadota:2016tqq}, while above the dashed line, millicharged DM would extract too much angular momentum from the Milky Way disk~\cite{Stebbins:2019xjr}.  Migdal effect constraints from SENSEI~\cite{SENSEI:2020dpa,Berghaus:2022pbu} and XENON10/XENON1T (\cite{Essig:2019xkx}, as shown in Ref.~\cite{Berghaus:2022pbu}) are shown in gray.}
    \label{fig:millicharge}
\end{figure}

Figure~\ref{fig:millicharge} shows, as a color scale, the DM charge corresponding to a given $m_{\chi}$ and $\sigma_{n\chi}$. In gray we superimpose the same Migdal effect limits shown in Fig.~\ref{fig:light}. In addition, we show two astrophysical bounds on millicharged DM, which are relevant in this parameter space as a result of the induced DM charge. First, Ref.~\cite{Stebbins:2019xjr} argued that fractionally charged DM interacting with Galactic magnetic fields in the Milky Way would extract angular momentum from the Milky Way disk, spinning down the disk over the course of gigayears. Although they report an order-of-magnitude uncertainty on their limit, Fig.~\ref{fig:millicharge} covers more than 10 orders of magnitude in $\epsilon$. Taking this uncertainty into account, these bounds still far supersede those set by the tree level interactions. Second, Ref.~\cite{Kadota:2016tqq} considered millicharged DM moving in galaxy clusters under the influence of cluster magnetic fields, and argued that if the DM charge were too large, magnetic fields would substantially alter the DM density profile. This results in another strong bound on DM charge, also shown in Fig.~\ref{fig:millicharge}. 

Other limits on millicharged DM may not apply, or may need to be considered more carefully in this scenario. For example, supernova cooling constraints on millicharged particles~\cite{Chang:2018rso}  do not apply here because the proton coupling is large enough to trap the DM within the proto-neutron star. Similarly, the argument that millicharged particles would be evacuated from the Galaxy by supernovae, put forward by Ref.~\cite{Chuzhoy:2008zy}, depends on the dark matter not scattering too frequently with Standard Model particles, an assumption that may be violated in at least some of the parameter space we consider. For this reason, and because the corresponding limit is weaker than the other astrophysical bounds we show, we do not plot the limit from Ref.~\cite{Chuzhoy:2008zy}. 
We also note that a specific model of DM that has a loop-induced effective electric charge through its hadronic couplings will interact with protons differently than a typical millicharged particle. This could even strengthen the astrophysical bounds shown in Fig.~\ref{fig:millicharge}, by, for example, enhancing the amount of angular momentum extracted from the Milky Way disk.

\section{Conclusions}\label{sec:conclusions}

We have presented a one-loop calculation of the low energy DM-electron cross section for DM which interacts exclusively with quarks at tree level.  This interaction should generically emerge in a wide range of DM models that interact with quarks through a vector mediator.  This has allowed us to derive novel constraints on the DM-proton cross section using existing constraints from SENSEI data.  We have shown that currently-running and upcoming electron recoil detectors, DAMIC-M and Oscura, should be able to probe DM-proton cross sections that may be beyond the reach of nuclear recoil detectors.  Finally we have demonstrated that DM that  interacts with quarks through a light mediator at tree level has an effective electric charge which can be used to recast astrophysical and cosmological constraints on the DM-electron cross section.   

Standard Model loop interactions can be an effective tool in exploring DM behavior, and are an inevitable but often-ignored part of any DM theory.  While we focused on quark scattering interactions through a vector mediator in this work, we note that loop interactions similar to those described in this work may be effective at bridging different DM-Standard Model interactions in annihilation processes and with mediators not explored here.

\begin{acknowledgments}
We are grateful to Kim Boddy, Humberto Gilmer, Martin Hoferichter, Matheus Hostert, Jason Kumar, David Morrissey, Marianne Moore, and Tim Tait for helpful discussions. This work is supported by the Natural Sciences and Engineering Research Council of Canada, the Arthur B. McDonald Canadian Astroparticle Research Institute, and the Canada Foundation for Innovation. Research at Perimeter Institute is supported by the Government of Canada through the Department of Innovation, Science, and Economic Development, and by the Province of Ontario.

\end{acknowledgments}
\onecolumngrid
\appendix
\section{Cross Section Calculation}
\label{sec:appendixA}
\subsection{Effective Lagrangians}

\begin{figure}[h]
    \centering
    \includegraphics[width=0.25\columnwidth]{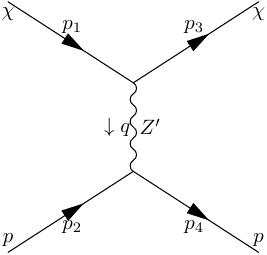}
    
    \caption{Tree level Feynman diagram for dark matter proton scattering}
    \label{fig:tree}
\end{figure}

\begin{figure}[h]
        \includegraphics[width=0.25\columnwidth]{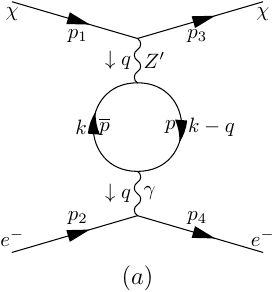}
        \includegraphics[width=0.25\columnwidth]{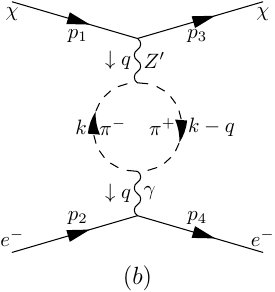}
        \includegraphics[width=0.25\columnwidth]{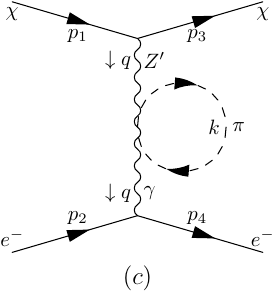}
    \caption{Dark matter-electron interaction induced by proton and pion loops.}
    \label{fig:loopsap}
\end{figure}

We calculate the dark matter (DM) proton cross section based on tree level interactions of the form shown in Fig.~\ref{fig:tree}  and the DM-electron scattering cross sections resulting from one loop diagrams of the form shown in Fig.~\ref{fig:loopsap}.  These calculations are all done in the center of mass frame, as the total cross sections are Lorentz invariant. We start with the proton tree level scattering.

We start the calculation by considering an underlying DM-quark interaction of the form
\begin{equation}
\label{eq:quarkcoupleap}\mathcal{L}\supset\sum_q\alpha_q Z'_{\mu}q\gamma^{\mu}\bar{q}+g_{\chi}Z'_{\mu}\chi\gamma^{\mu}\bar{\chi}
\end{equation}
where $\chi$ is the massive DM particle, $q$ are the quarks, $\alpha_q$ is the coupling of each quark specie to $Z'_{\mu}$, $g_{\chi}$ is the coupling of $\chi$ to its vector mediator.   In the low energy limit quarks are confined to light mesons and baryons. Their behavior is best described my Chiral Effective Field Theory (ChEFT) with Baryons.  At low energies \eqref{eq:quarkcoupleap} gives an effective coupling to protons of the form \cite{Bishara_2017}
\begin{equation}
    \bra{P'}\sum_q\alpha_q Z'_{\mu}q\gamma^{\mu}\bar{q}\ket{P}=(2\alpha_u+\alpha_d)Z'_{\mu}P'\gamma^{\mu}P.
\end{equation}
  
The lowest order terms in the ChEFT Lagrangian that will contribute to the meson-$Z'$ interaction  are
\cite{Ecker:1988te}
\begin{equation}
\label{eq:chiralLap}
    L\supset \frac{F^2}{4}Tr\left(D_{\mu}U D^{\mu}U^{\dag}\right),
\end{equation}
where $U= e^{\frac{i}{F}\pi}$ contains the light meson octet 
\begin{equation}
    \pi= 
    \begin{pmatrix}
    \pi^0+\frac{\eta_8}{\sqrt{3}}&\sqrt{2}\pi^+ &\sqrt{2}K^+ \\ \sqrt{2}\pi^-&-\pi^0+\frac{\eta_8}{\sqrt{3}} &\sqrt{2}K^0\\
    \sqrt{2}K^-&\sqrt{2}\bar{K}^0 & -\frac{2}{\sqrt{3}}\eta_8
    \end{pmatrix}.
    \end{equation}
Here $F$ represents the pion decay constant.  Interactions with external vector fields at highest order are captured in the derivative terms
\begin{equation}
\begin{split}
    &D_{\mu}U =\partial_{\mu}U-iv_{\mu} U+iUv_{\mu},\\
    &D_{\mu}U^{\dag}=\partial_{\mu}U^{\dag}+iU^{\dag}v_{\mu}-iv_{\mu} U^{\dag},\\
\end{split}
\end{equation}
where
$v_{\mu}$ represents an interaction with an external vector.  To capture the interaction with $Z'$ depicted in \eqref{eq:quarkcoupleap}, we take $v_{\mu}=Z'_{\mu}\text{diag}(\alpha_u,\alpha_d,\alpha_s)$.  We can include electromagnetic interactions here as well by including a term of the form $v_{\mu} = e A_{\mu}\text{diag}(\frac{2}{3},-\frac{1}{3},-\frac{1}{3})$.  
Expanding out the chiral lagrangian \eqref{eq:chiralLap} gives the following interaction terms between light mesons, photons, and $Z'$:
\begin{equation}
\begin{split}
    \mathcal{L}\supset &i(\alpha_u-\alpha_d)Z'_{\mu}(\pi^-\partial^{\mu}\pi^+ - \pi^+\partial^\mu\pi^-)\\+&i(\alpha_u-\alpha_s)Z'_{\mu}(K^-\partial^{\mu}K^+ - K^+\partial^{\mu} K^-)\\
    +&ieA_{\mu}(\pi^-\partial^{\mu}\pi^+ - \pi^+\partial^\mu\pi^-)\\+&ieA_{\mu}(K^-\partial^{\mu}K^+ - K^+\partial^{\mu} K^-)\\
    +&2 e(\alpha_u-\alpha_d)Z'_{\mu}A^{\mu}\pi^+\pi^-\\
    +&2 e(\alpha_u-\alpha_s)Z'_{\mu}A^{\mu}K^+K^-~.\\
\end{split}
\end{equation}
\subsection{Tree Level Cross Section}
Now these effective interaction terms can be used to calculate the cross sections resulting from Fig.~\ref{fig:tree}
 and Fig.~\ref{fig:loopsap}
The amplitude of the tree-level proton scattering diagram is
\begin{equation}
    \mathcal{M}_{\text{tree}}=g_{\chi}(2\alpha_u+\alpha_d)u_{\chi}(p_1)\gamma^{\mu}\Bar{u}_{\chi}(p_3)\frac{g_{\mu \nu}}{m_{Z'}^2-(p_1-p_3)^2}u_P(p_2)\gamma^{\nu}\Bar{u}_P(p_4).
\end{equation}
Squaring this and summing over polarisation gives a cross section of
\begin{equation}
\label{protontreeap}
    \frac{d\sigma_{\chi p}}{d\Omega} = \frac{g_{\chi}^2 (2\alpha_u+\alpha_d)^2 \mu_{\chi p}^2}{4\pi^2 (m_Z^2-t)^2},
\end{equation}
where $m_{Z'}$ is the mediator mass respectively.  We use $\mu_{a b} = \frac{m_a m_b}{m_a+m_b}$ to represent the reduced mass of two particles $a$ and $b$ with masses $m_a$ and $m_b$.  Above $\mu_{\chi p}^2$ represents the reduced mass of $\chi$ and the proton.  We take $m_p = 0.938$ GeV. We note that many model specific terms such as $g_{\chi}$ and $m_{Z'}$ will divide out in the ratio between the tree level and loop level cross section.  
\subsection{Loop Level Calculations}
\subsubsection{Proton Loop}
\label{sec:protonloop}
Calculating the DM-electron scattering cross section will require a few additional steps compared to the DM-proton cross section.  We first calculate the  loop terms and then calculate their contribution to the overall scattering cross section.  

The proton loop piece of the diagram shown in Fig.~\ref{fig:tree} (a) can be derived by modifying the calculation for QED vacuum polarization found in Ref.~\cite{Shwartz}.  The loop contribution takes the form 
\begin{equation}
    i\Pi_{\mu \nu}^P= -e(2\alpha_u+\alpha_d)\int\frac{d^4k}{(2\pi)^4}\frac{Tr[\gamma^{\mu}(\cancel{k}-\cancel{q}+m_P)\gamma^{\nu}(\cancel{k}+m_p)]}{((q-k)^2-m_P^2)(k^2-m_P^2)}\,,
\end{equation}
where $q=p1-p3$. Taking the trace gives
\begin{equation}
    i\Pi_{\mu \nu}^P= -4e(2\alpha_u+\alpha_d)\int\frac{d^4k}{(2\pi)^4}\frac{2k^{\mu}k^{\nu}+g^{\mu \nu}(-k^2+q\cdot k+m_P^2)}{((q-k)^2-m_P^2)(k^2-m_P^2)}\,.
\end{equation}
Here we have dropped $q^{\mu}q^{\nu}$ terms as these do not contribute to our final cross section.  Using the relation 
$\frac{1}{AB}=\int_0^1 dx\frac{1}{(A+(B-A)x)^2}$,  shifting $k^{\mu}$ to $k^{\mu} +q^{\mu}(1-x)$,  and dropping the $q^{\mu}q^{\nu}$ terms and terms odd in $k$ from the numerator (these will go to zero or not contribute to the final cross section) give 
\begin{equation}
    \Pi_{\mu \nu}^P= 4ie(2\alpha_u+\alpha_d)\int\frac{d^4k}{(2\pi)^4}\int_0^1 dx\frac{2k^{\mu}k^{\nu}-g^{\mu\nu}(k^2-x(1-x)q^2-m_P^2)}{(k^2+q^2x(1-x)-m_P^2)^2}\,.
\end{equation}
This integral will diverge in 4 dimensions.  We manage this divergence using dimensional regularization by integrating over 
$d$ dimensions then taking the limit $d\rightarrow4-\epsilon$.  To do these integrals, we take $k^{\mu\nu} \rightarrow \frac{1}{d}k^2g^{\mu\nu}$.  In $d$ dimensions the integral takes the form
\begin{equation}
\label{lastint}
    \Pi_{\mu \nu}^P= -4ie(2\alpha_u+\alpha_d)\mu^{d-4}g^{\mu\nu}\int\frac{d^dk}{(2\pi)^d}\int_0^1 dx\frac{((1-\frac{2}{d})k^2-x(1-x)q^2-m_P^2)}{(k^2+q^2x(1-x)-m_P^2)^2}\,.
\end{equation}
Using the following formulas
\begin{equation}
    \begin{split}
        &\int \frac{d^dk}{(2\pi)^d}\frac{k^2}{(k^2-\Delta+i\epsilon)^2} = -\frac{d}{2}\frac{i}{(4\pi)^{d/2}}\frac{1}{\Delta^{1-\frac{d}{2}}}\Gamma\left(\frac{2-d}{2}\right)\\
        &\int \frac{d^dk}{(2\pi)^d}\frac{1}{(k^2-\Delta+i\epsilon)^2} = \frac{i}{(4\pi)^{d/2}}\frac{1}{\Delta^{2-\frac{d}{2}}}\Gamma\left(\frac{4-d}{2}\right)\,.
    \end{split}
\end{equation}
we complete the $k$ integral in Eq.\eqref{lastint}.  
This gives
\begin{equation}
    \Pi_{\mu \nu}^P= -8e(2\alpha_u+\alpha_d)\mu^{4-d}g^{\mu\nu}q^2\Gamma\left(2-\frac{d}{2}\right)\int_0^1 dx x(1-x)\frac{1}{(m_P^2-q^2x(1-x))^{2-\frac{d}{2}}}\,.
\end{equation}
If we take the limit $p\ll m_P$ then the integral over $x$ is simple and gives
\begin{equation}
    \Pi_{\mu \nu}^P= -\frac{e(2\alpha_u+\alpha_d)}{12\pi^2}g^{\mu\nu}q^2\left(\frac{\mu}{m_P}\right)^{4-d}\Gamma\left(2-\frac{d}{2}\right)\,.
\end{equation}
This diverges in the $\epsilon=0$ limit, expanding around $\epsilon=0$ gives
\begin{equation}
    \Pi_{\mu \nu}^P= -\frac{e(2\alpha_u+\alpha_d)}{24\pi^2}g^{\mu\nu}q^2\left(\frac{2}{\epsilon}+\text{Log}\left[\frac{4\pi e^{\gamma_E}\mu^2}{m_P^2}\right]\right)\,.
\end{equation}
Keeping only the finite terms gives proton loop contribution to the scattering diagram
\begin{equation}
\label{eq:protonloopap}
    \Pi_{\mu \nu}^P= -\frac{e(2\alpha_u+\alpha_d)}{24\pi^2}g^{\mu\nu}q^2\text{Log}\left[\frac{4\pi e^{\gamma_E}\mu^2}{m_P^2}\right]\,.
\end{equation}
Here we define $\mu$ as a cutoff scale for low energy ChEFT with Baryons.  Following Ref.~\cite{Becher:1999he} we take $\mu=m_P$.

\subsubsection{Light Meson Loops}
Both pions and kaons can contribute to DM-electron loop level scattering.  There are two different loop diagrams for each light meson particle.  The general form of these diagrams is displayed in Fig.~\ref{fig:loopsap}  (b) and (c). Below, we calculate a general expression for these loop contributions, and then add in the specific coupling and mass values for specific mesons.  These calculations modify the vacuum polarization derivations for scalar QED shown in \cite{Shwartz}.  Using the interaction terms of the lagrangian shown in Eq.\eqref{eq:chiralL} we get the following amplitudes for the loop in Fig.~\ref{fig:loopsap}(b) and Fig.~\ref{fig:loops} (c) respectively

\begin{equation}
\begin{split}
      i\Pi^b_{\mu\nu} & = -e g\int \frac{d^4k}{(2\pi)^4}\frac{i(2k^{\mu}-q^{\mu})}{(k-q)^2-m^2+i\epsilon}\frac{i(2k^{\mu}-q^{\mu})}{k^2-m^2+i\epsilon}, \\
      i\Pi^c_{\mu\nu} & = -e g\int \frac{d^4k}{(2\pi)^4}\frac{i}{k^2-m^2+i\epsilon}\,.
\end{split}
\end{equation}
We use $g$ as a stand in for the meson-$Z$ coupling and $m$ to represent the meson mass. Summing these together and then preforming the same integration, approximation and simplification steps outlined in Section \ref{sec:protonloop} gives the following total loop contribution 
\begin{equation}
    \Pi_{\mu\nu}^{\text{meson}}=-\frac{eg}{48\pi^2}g^{\mu\nu}q^2Log\left[\frac{4\pi e^{-\gamma_E}\mu^2}{m^2}\right]\,.
\end{equation}
The specific loop contributions from pions and kaons are then
\begin{equation}
\label{eq:pionloopap}
    \begin{split}
\Pi_{\mu\nu}^{\pi}&=-\frac{e(\alpha_u-\alpha_d)}{48\pi^2}g^{\mu\nu}q^2Log\left[\frac{4\pi e^{-\gamma_E}\mu^2}{m_{\pi}^2}\right]\\
        \Pi_{\mu\nu}^{K}&=-\frac{e(\alpha_u-\alpha_s)}{48\pi^2}g^{\mu\nu}q^2Log\left[\frac{4\pi e^{-\gamma_E}\mu^2}{m_{K}^2}\right]\,.
\end{split}
\end{equation}
\subsubsection{Loop Level Cross Section}
Combining the loop contributions calculated in \eqref{eq:protonloopap} and \eqref{eq:pionloopap} gives the overall loop contribution at low energies
\begin{equation}
    \Pi_{\mu\nu}^{\text{loop}}= -\frac{e}{48\pi^2}g^{\mu\nu}q^2 c_{\text{
    loop
    }}\,,
\end{equation}
where 
\begin{equation}
    c_{\text{loop}} =  -4(2\alpha_u+\alpha_d)\text{Log}\left(\frac{4 \pi e^{-\gamma_E}\mu^2}{m_{p}^2}\right)
    -(\alpha_u-\alpha_d)\text{Log}\left(\frac{4 \pi e^{-\gamma_E}\mu^2}{m_{\pi}^2}\right)
    -(\alpha_u-\alpha_s)\text{Log}\left(\frac{4 \pi e^{-\gamma_E}\mu^2}{m_{K}^2}\right)\,.  
\end{equation}
The amplitude of the full DM-electron scattering interaction is
\begin{equation}
    \mathcal{M}_{\text{loop}}=g_{\chi}eu_{\chi}(p_1)\gamma^{\mu}\Bar{u}_{\chi}(p_3)\frac{g_{\mu \nu}}{m_{Z'}^2-(p_1-p_3)^2}\Pi_{\nu\omega}^{\text{loop}}\frac{g_{\omega \xi}}{(p_1-p_3)^2}u_e(p_2)\gamma^{\xi}\Bar{u}_e(p_4)\,.
\end{equation}
This gives a final cross section of
\begin{equation}
    \frac{d\sigma_{\chi e}}{d\Omega} =
   \frac{g_{\chi}^2e^2\mu_{\chi e}^2}{144(4\pi)^6(m_{Z'}^2-t)^2} c_{loop}^2\,,
\end{equation}
\begin{equation}
    \frac{d\sigma_{\chi e}}{d\Omega} =
   \frac{d\sigma_{\chi p}}{d\Omega} \frac{e^2}{2304 \pi^4 (2 \alpha_u+\alpha_d)^2}  \left(\frac{\mu_{\chi e}}{\mu_{\chi p}}\right)^2 c_{loop}^2\,.
\end{equation}

For a heavy mediator, we can perform the integration explicitly to relate the total cross sections:

\begin{equation}
    \sigma_{\chi e} =
   \sigma_{\chi p} \frac{e^2}{2304 \pi^4 (2 \alpha_u+\alpha_d)^2}  \left(\frac{\mu_{\chi e}}{\mu_{\chi p}}\right)^2 c_{loop}^2\,.
\end{equation}

In the case of a light mediator, the total cross section diverges. Literature involving light mediators often parameterizes the differential cross section in terms of a  reference cross section $\overline{\sigma}$:

\begin{equation}
    \frac{d\sigma_{\chi T}}{dq^2} = \frac{\overline{\sigma}_{\chi T}}{4\mu_{\chi T}^2v^2} \left(\frac{q_{ref}}{q}\right)^4\,,
\end{equation}
where $q_{ref}$ is a reference momentum, usually taken to be $\sim\alpha m_e$~\cite{Emken:2019tni}. 
We can rewrite our differential cross sections in terms of the momentum transfer $q$, to match direct detection literature, by noting that 

\begin{equation}
    q^2 = \bf{p_1} + \bf{p_3} - 2\bf{p_1}\bf{p_3}\,,
\end{equation}
so
\begin{equation}
    \frac{d\sigma}{d\cos{\theta}} = 2\mu_{\chi T}^2v^2\frac{d\sigma}{dq^2}\,
\end{equation}
where $T$ denotes the target of the scattering, either proton or electron. We can thus write

\begin{equation}
    \frac{d\sigma_{\chi e}}{dq^2} =
   \frac{d\sigma_{\chi p}}{dq^2} \frac{\mu_{\chi p}^2}{\mu_{\chi e}^2} \frac{e^2}{2304 \pi^4 (2 \alpha_u+\alpha_d)^2}  \left(\frac{\mu_{\chi e}}{\mu_{\chi p}}\right)^2 c_{loop}^2\,,
\end{equation}
and given the parametrization above, we can relate the reference cross sections:

\begin{equation}
    \overline{\sigma}_{\chi e} =
   \overline{\sigma}_{\chi p} \frac{e^2}{2304 \pi^4 (2 \alpha_u+\alpha_d)^2}  \left(\frac{\mu_{\chi e}}{\mu_{\chi p}}\right)^2 c_{loop}^2\,.
\end{equation}
As it turns out, the relation between the reference cross sections is exactly the relation between the total cross sections found in the heavy mediator case.

\section{Proton-Only Loops}
\label{sec:appendixB}
In the main text, we considered the case where $\alpha_d = -\alpha_u$, and $\alpha_i = 0$ for all other quarks. We can also consider the case where the couplings to all (light) quarks are equal, i.e. $\alpha_u = \alpha_d = \alpha_s$. In this case, only the proton contributes, resulting in limits that are weaker by a factor of 7---8. We show the results below.

\begin{figure}[h!]
    \centering
    \includegraphics[width=\columnwidth]{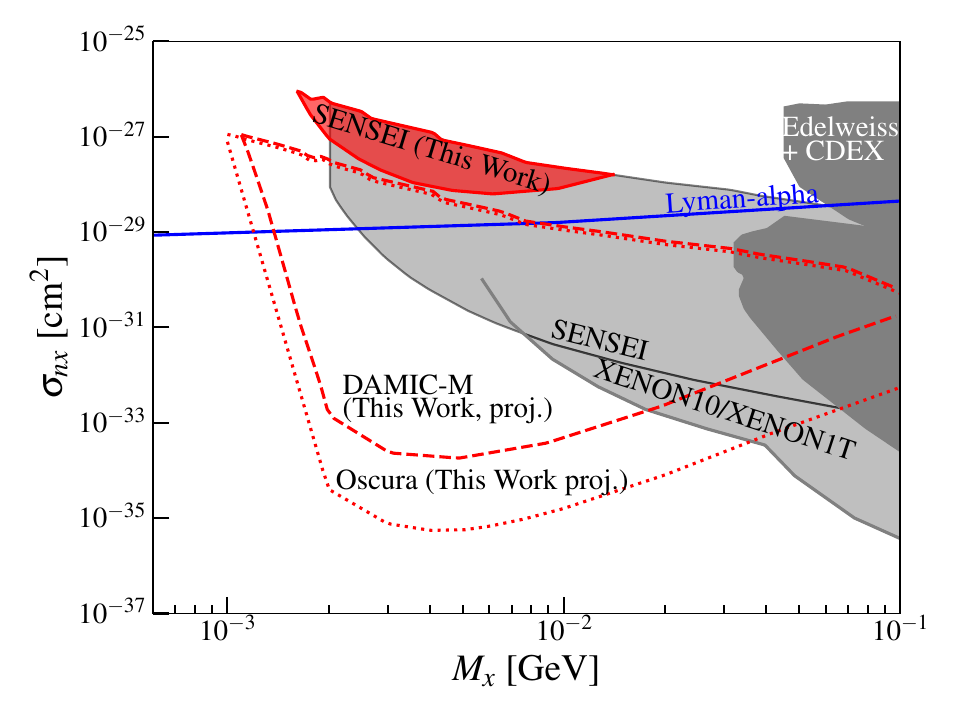}
    
    \caption{Limits on DM with a heavy vector mediator and $\alpha_u=\alpha_d=\alpha_s$.  Our recasting of constraints from SENSEI ~\cite{SENSEI:2020dpa} is shown in red, while the regions outlined in red dashed and dotted lines will be accessible to DAMIC-M ~\cite{Battaglieri:2017aum,Settimo:2018qcm,Settimo:2020cbq} and Oscura ~\cite{2022arXiv220210518A} respectively.  Existing detector constrains From Migdal effect searches at SENSEI \cite{SENSEI:2020dpa,Berghaus:2022pbu}, XENON10/1T  (\cite{Essig:2019xkx}, as shown as in Ref.~\cite{Berghaus:2022pbu}), CDEX \cite{CDEX:2019hzn,CDEX:2021cll} and EDELWEISS \cite{EDELWEISS:2019vjv,2022PhRvD.106f2004A} are shown in gray, while Lyman alpha constraints \cite{Rogers:2021byl} are shown in blue.}
    \label{fig:heavyproton}
\end{figure}

\begin{figure}
    \centering
    \includegraphics[width=\columnwidth]{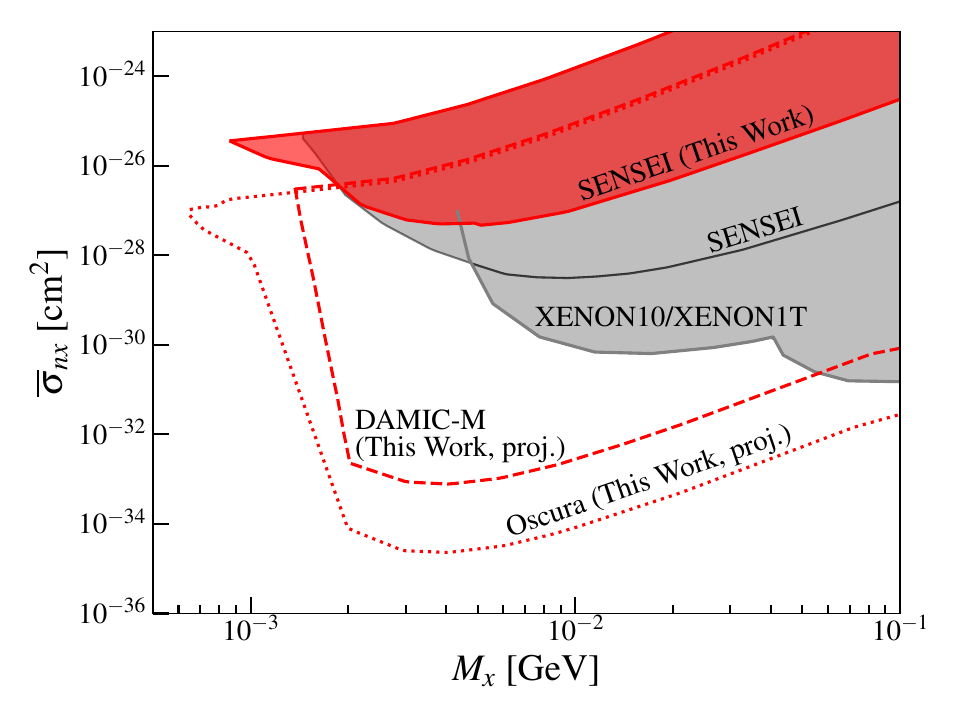}
    
    \caption{Limits on DM scattering through a light vector mediator and $\alpha_u=\alpha_d=\alpha_s$. Our results for SENSEI ~\cite{SENSEI:2020dpa}  are highlighted in red, and the expected reach of DAMIC-M ~\cite{Battaglieri:2017aum,Settimo:2018qcm,Settimo:2020cbq} and Oscura ~\cite{2022arXiv220210518A} are outlined by the red dashed and dotted lines respectively. Constraints from Migdal effect searches at SENSEI \cite{SENSEI:2020dpa,Berghaus:2022pbu}, and XENON10/1T (\cite{Essig:2019xkx}, as shown in Ref.~\cite{Berghaus:2022pbu}) are shown in gray.}
    \label{fig:protonlight}
\end{figure}

\begin{figure}
    \centering
    \includegraphics[width=\columnwidth]{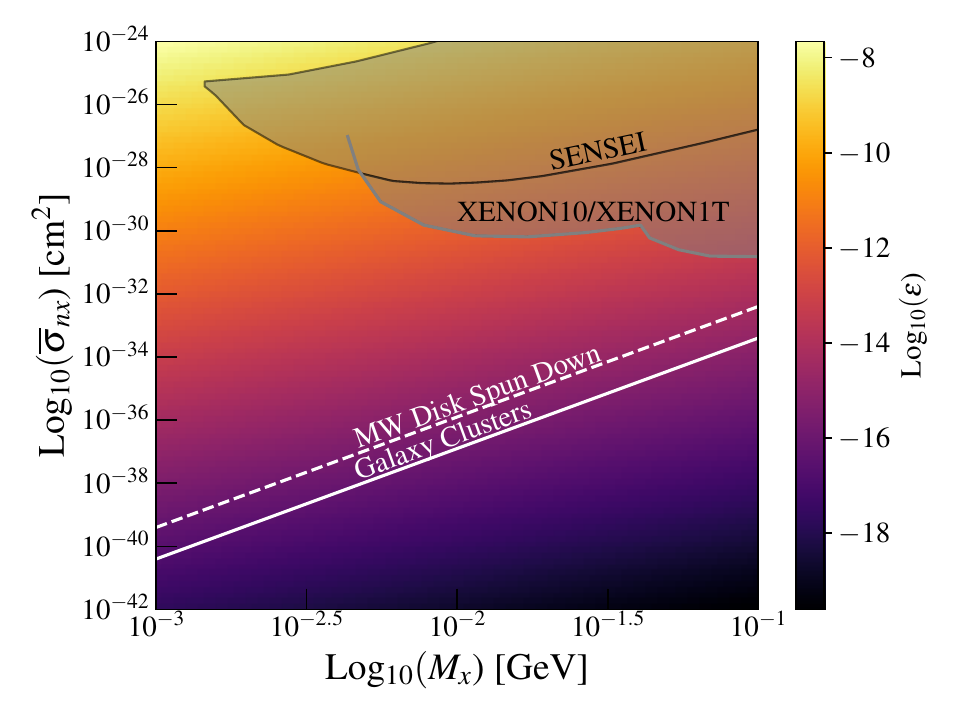}
    
    \caption{Effective millicharge resulting from hadronic loop interactions assuming $\alpha_u=\alpha_d=\alpha_s$. The white lines are astrophysical constaints on millicharged DM: above the solid line, ccluster magnetic fields would noticeably alter the density profile of galaxy clusters~\cite{Kadota:2016tqq}, while above the dashed line, millicharged DM would extract too much angular momentum from the Milky Way disk~\cite{Stebbins:2019xjr}.  Migdal effect constraints from SENSEI~\cite{SENSEI:2020dpa,Berghaus:2022pbu} and XENON10/XENON1T (\cite{Essig:2019xkx}, as shown in Ref.~\cite{Berghaus:2022pbu}) are shown in gray.}
    \label{fig:millichargeap}
\end{figure}

\clearpage

\bibliography{main}


\end{document}